\documentclass[pra,twocolumn,showpacs,superscriptaddress]{revtex4}
\usepackage{amsmath, float} 
\usepackage{amsfonts}
\usepackage{amssymb}
\usepackage{graphicx}
\usepackage{bm}
\usepackage{enumerate}
\usepackage{color}
\usepackage{amsthm}

\theoremstyle{plain}
\newtheorem{theorem}{Theorem}
\newtheorem{Proposition}[theorem]{Proposition}

\newtheorem{definition}[theorem]{Definition}

\theoremstyle{remark}
\newtheorem{remark}{Remark}

\begin{document}

\title{Multi-particle decoherence free subspaces in extended systems}
\author{Raisa I. Karasik}
\affiliation{Applied Science \& Technology, 
        University of California, Berkeley, California 94720, USA}
\affiliation{Department of Chemistry and Berkeley Quantum Information Center, 
        University of California, Berkeley, California 94720, USA}
\affiliation{Institute for Quantum Information Science,
        University of Calgary, Calgary, Alberta T2N 1N4, Canada}
\author{Karl-Peter Marzlin}
\affiliation{Institute for Quantum Information Science,
        University of Calgary, Calgary, Alberta T2N 1N4, Canada}
\author{Barry C.~Sanders}
\affiliation{Institute for Quantum Information Science,
        University of Calgary, Calgary, Alberta T2N 1N4, Canada}
\author{K. Birgitta Whaley}
\affiliation{Applied Science \& Technology, 
        University of California, Berkeley, California 94720, USA}
\affiliation{Department of Chemistry and Berkeley Quantum Information Center, 
        University of California, Berkeley, California 94720, USA}        
\begin{abstract}
We develop a method to determine spatial configurations to realize decoherence-free subspaces for spatially extended multi-particle systems. We have assumed normal reservoir behavior including translational invariance of the reservoir and preparation in stationary states or mixture thereof and weak Markovian system-reservoir coupling that requires energy transfer. One important outcome of our method is a proof that there does not exist a multi- 
particle decoherence-free subspace in such systems except in the limit that the spatial extent of the system becomes infinitesimal.
\end{abstract}

\maketitle
\section{Introduction}
\label{sec:intro}

The notion of a decoherence-free (or noiseless) subspace (DFS) has been proposed as one strategy
to combat the effects of decoherence in quantum computation~\cite{Duan, Zanardi, Lid98, Zan98} and 
has the advantage of being passive, in contrast to quantum error correction, 
which requires active syndrome measurement and feed forward~\cite{Shor, Gott, Knill}. The construction of a DFS relies
on the existence of symmetries in the decoherence process that allow certain subspaces of
quantum states to be completely decoupled from the environment. A DFS is of practical
interest because it would reduce errors due to decoherence, thereby lowering the
quantum error correction overhead in quantum information processing, and be especially useful
in providing longer quantum memory.

Extensive formal analyses of the DFS have been performed~
\cite{Kempe,Shabani,Lid99,Wu,Dua98,Kri05,Kri06}
and DFS have been applied in quantum information processing  
to reduce errors due to decoherence \cite{Kwiat,Kie01,Vio01, Moh03, Lan05}.
However, realizing
a DFS (or approximate DFSs that would reduce decoherence over a finite, but long time
scale) under general conditions is required. Our goal here is to
address this need, namely to introduce general realistic principles for
decoherence of  systems of particles, to create a formalism for describing these
systems in order to determine systematically when a perfect DFS is impossible, and to demonstrate by
example how an approximate DFS can be created even though a perfect DFS cannot
be.

We assume that the DFS is encoded into a collection of~$N$ identical particles, each
with a $D$-dimensional internal Hilbert space ${\cal H}_n$ spanned by the states
\begin{equation}
        \{|\alpha\rangle_n;\alpha=1,\ldots,D\} \; , \; n=1,\ldots,N
\end{equation}
Although the particles are identical, they are not co-located (except
when we refer to the ``Dicke Limit''~\cite{Dic54} in which the separation of particles is allowed
to tend to zero). The position of particle $n$ is denoted $\mathbf{x}_n$.
In the absence of any coupling to the environment, each internal level $|\alpha\rangle_n$
is a stationary state of the system's Hamiltonian evolution with eigenvalue~$\omega_\alpha$
(we set $\hbar=1$).

This system interacts with a reservoir, which is assumed to have no memory (Markovian).
At time~$t=0$, the state of the system~$\rho_\text{S}$ and the state of the reservoir~$\rho_\text{R}$ are independent and then commence interacting. This separability of system and reservoir states is standard in the theory of open quantum systems, and we assume that the evolution of the system in the presence of the reservoir can be described by a completely positive map on the system, or equivalently by a Lindblad master equation~\cite{Gor76, Lindblad}. We add the additional assumption that the reservoir is translationally invariant, which is reasonable but is also an important and restrictive assumption in studying spatially extended systems of particles. The generator of translation is a vector momentum operator $\hat{\mathbf{P}}$. We assume that the Hamiltonian reservoir dynamics depend on $|\hat{\mathbf{P}}|$ and not on the direction of $\hat{\mathbf{P}}$, i.e. the reservoir has no preferred direction. We refer to a reservoir satisfying these conditions as a homogeneous isotropic Markovian reservoir. 

Finally we place some reasonable restrictions on the interaction between system and reservoir. Specifically, 
we consider an interaction Hamiltonian of the form
\begin{equation}  
  H_I = -\sum_{n=1}^N \sum_i \hat{d}_n^i \hat{E}^i(\mathbf{x}_n),
\label{hInt}
\end{equation}
with reservoir operators $\hat{E}^i(\mathbf{x}_n)$, which depend on the position
of the interacting particle $n$, and system operators
\begin{equation}
        \hat{d}_n^{i}\equiv\!\!\sum_{\stackrel{\alpha,\, \beta;}{ 
        \omega_{\alpha\beta}\neq 0}}\!\!d_{\alpha\beta}^{i}\hat{\sigma}_{n,\alpha\beta},\;
        \hat{\sigma}_{n,\alpha\beta}\equiv|\alpha\rangle_n\langle\beta|
\label{eq:dn}
\end{equation}
that raise and lower the system energy levels. Here
$\omega_{\alpha\beta}\equiv\omega_{\alpha}-\omega_{\beta}\neq 0$ in the sum,
which ensures that terms involving states $|\alpha\rangle$  and
$|\beta\rangle$ of the same energy are excluded. Thus the reservoir absorbs
and delivers energy
to the system but does not contribute to strict dephasing. 
The Hamiltonian (\ref{hInt}) will be used to derive a master equation for the system.

In summary, we have a system of particles with internal levels interacting linearly with a reservoir
that is translationally invariant, whose Hamiltonian is independent of direction. These assumptions
describe typical realistic systems, for example atomic or molecular gases, but have serious
implications for constructing a DFS. On the one hand, we will find that we rule out the existence of a DFS under
quite general conditions, but on the other hand, we will establish a foundation for obtaining a
decoherence-suppressed subspace that is an approximate DFS.

\section{The decoherence-free subspace}

The density operator for the system, traced over the reservoir, is given by $\rho$, which is a bounded
operator on the Hilbert space~$\mathcal{H}_\text{S}$ for the system of~$N$ particles, each with~$D$ internal
levels. Under the assumptions given in Sec.~\ref{sec:intro}, 
we can express the evolution of~$\rho$ by the master equation
\begin{equation} 
        \dot{\rho} =  -i[\hat{H}_\text{eff},\rho]+L_D[\rho],
\label{eq:master}
\end{equation}
with 
\begin{equation}
        \hat{H}_\text{eff}=\hat{H}_\text{S}+\hat{\Delta}
\label{eq:heff}
\end{equation}
the Hermitian effective system Hamiltonian, where $\hat{H}_\text{S}$
corresponds to the Hermitian system Hamiltonian, $\hat{\Delta}$ is the
Hermitian contribution from the environment, and the nonunitary dynamics are 
incorporated into the decoherence propagator
\begin{equation}
\label{eq:LD}
        L_D[\rho]=\frac{1}{2}\sum_{l=1}^M\lambda_l \left([\hat{J}_l,\rho
        \hat{J}_l^\dagger]+[\hat{J}_l\rho, \hat{J}_l^\dagger]\right).
\end{equation}
In Eq.~(\ref{eq:LD}), the real numbers $\lambda_l>0$ are time-independent and operators
$\{\hat{J}_l\}_{l=1}^{M}$ form a subset of a complete basis for the space of bounded operators defined on~$\mathcal{H}_\text{S}$.

We use the same definition for DFS as in \cite{Kar}.
Let $D(\mathcal{H})$ be the set of all density matrices defined for a quantum
system associated with the Hilbert space $\mathcal{H}$.
\begin{definition}
Let the time evolution of an open quantum system with Hilbert space
${\cal H}_\text{S}$ be governed by Eq.~(\ref{eq:master}).
Then a decoherence free subspace $\mathcal{H}_\text{DFS}$ is a subspace of $\mathcal{H}_\text{S}$ such that all density matrices $\rho(t)\in D(\mathcal{H}_\text{DFS})$  fulfill
\begin{equation}
 \dot{\rho} =  -i[\hat{H}_\emph{eff},\rho] \; \forall t \; .
\end{equation}
\label{def:DFS}
\end{definition}
\begin{remark}
Note that Definition \ref{def:DFS} implies that $L_D[\rho]=0$ for all times $t$. 
\end{remark}
\begin{remark}
        The evolution of a state in $\mathcal{H}_\text{DFS}$ includes the
        environment-induced unitary evolution given by $\hat{\Delta}$, but not the decoherence effects.
\end{remark}
\begin{remark}
\label{alteDFS}
        An alternative definition for a DFS requires that states experience unitary evolution 
        due to the system Hamiltonian $\hat{H}_\text{S}$ only. In this case states are unchanged by 
        (unitary and non-unitary) interactions with the environment~\cite{Zan98, Zan97}
        (i.e.~$\rho(t)=\text{e}^{-i t\hat{H}_\text{S}}\rho\text{e}^{i t\hat{H}_\text{S}}$ or, 
        equivalently $\dot{\rho}= -i[\hat{H}_\text{S}, \rho]$ and 
         $-i[\hat{\Delta},\rho]+L_D[\rho]=0$). In Appendix~\ref{app:defDFS}, we show that for the 
        systems we consider, this 
        definition is a special restricted case of Definition~\ref{def:DFS}.
\end{remark}
To determine necessary conditions for DFS we introduce
\begin{definition}
An instantaneous pure decoherence-free subspace (IPDFS) $\mathcal{H}_\text{IPDFS}$ is a subspace of $\mathcal{H}_\text{S}$  such that all density matrices $\rho\in D(\mathcal{H}_\text{IPDFS})$ fulfill $\rho^2=\rho$ and $L_D[\rho]=0$.
\label{def:ipdfs}
\end{definition}
\begin{remark}
We note that in general condition $L_D[\rho(t)]=0$ at a specific time $t$ does not guarantee unitary evolution of $\rho(t)$ at a later time because $\hat{H}_\text{eff}$ can potentially drive $\rho(t)$ out of IPDFS. However, every state in DFS is an instantaneous pure decoherence-free state. Thus DFS is a subset of IPDFS.
\end{remark}

Now we establish a simple criterion for identifying whether a state is in an  IPDFS and thus determine necessary condition for DFS, beginning with a definition of the decoherence operator.
\begin{definition}
        The decoherence operator for a system with decoherence propagator~(\ref{eq:LD}) is
        \begin{equation}
        \label{eq:decoherencematrix}
                \hat{\Gamma} \equiv \sum_{l=1}^M\lambda_l\hat{J}^{\dagger}_l\hat{J}_l.
        \end{equation}
\end{definition}
\begin{remark}
        Note that the operator $\hat{\Gamma}$ is positive semidefinite because 
        $\forall$ $|\psi\rangle\in\mathcal{H}_\text{S}$,
        $\langle\psi|\hat{\Gamma}|\psi\rangle =
        \sum_{l=1}^M\lambda_l\left\|\hat{J}_l |\psi \rangle \right\|^2\geq 0$.
\end{remark}
Inserting expression~(\ref{eq:heff}) for $\hat{H}_\text{eff}$ into
Eq.~(\ref{eq:master}) and expression~(\ref{eq:decoherencematrix}) for the
decoherence operator into Eq.~(\ref{eq:LD}), we see 
that general form of the master equation can be written as
\begin{align}
        \dot{\rho}=&-i[\hat{H}_\text{S},\rho]
        +(-i\hat{\Delta}-\frac{\hat{\Gamma}}{2})\rho+
        \rho(i\hat{\Delta}-\frac{\hat{\Gamma}}{2})\nonumber \\&
        +\sum_{l=1}^M \lambda_l \hat{J}_l \rho\hat{J}_l^\dagger.
\label{Eq:generalME}
\end{align}
Henceforth we refer to the term $i\hat{\Delta}+\frac{1}{2} \hat{\Gamma}$ as the non-Hermitian transition operator.

The following proposition establishes necessary and sufficient conditions for the existence of an instantaneous pure decoherence-free state. A full proof is given in~\cite{Kar}. 
\begin{Proposition}
\label{prop:DFS0}
 For a pure state $\rho=|\psi\rangle\langle\psi|$, $L_D[\rho]=0$ iff
        $\hat{J}_l|\psi\rangle=c_l|\psi\rangle$ and 
        $\hat{\Gamma}|\psi\rangle=g|\psi\rangle$, where 
        $g=\sum_{l=1}^M \lambda_l |c_l|^2$.
\end{Proposition}

We demonstrate later that only operators $\hat{J}_l$ that have no non-zero eigenvalues appear in the systems we consider in this paper. In this case the following proposition becomes very useful.

\begin{Proposition}
\label {prop:DFS}
        Suppose that all eigenvalues of operators $\hat{J}_l$  for $l=1,\ldots, M$ 
        are equal to zero; then the state  $|\psi\rangle$ is in  an IPDFS  iff
        it is an eigenvector of the 
        decoherence operator $\hat{\Gamma}$~(\ref{eq:decoherencematrix}) with an eigenvalue of zero.
\end{Proposition}
\begin{proof}
        From Proposition~\ref{prop:DFS0} we know  
        that if $|\psi\rangle$ is in an IPDFS then
        $\hat{J}_l|\psi\rangle= c_l\, |\psi\rangle$ for  $l=1,\ldots, M$. The	
        supposition for Proposition~\ref{prop:DFS} implies 
        that $c_l=0$ for $l=1,\ldots, M$.
        Thus
        \begin{equation}
          \hat{\Gamma}|\psi\rangle=
          \sum_{l=1}^M\lambda_l\hat{J}_l^\dagger\hat{J}_l|\psi\rangle=0.
        \end{equation}
        Conversely, if 
        \begin{equation}
                \hat{\Gamma}|\psi\rangle=0|\psi\rangle,
        \end{equation}
        (with $\left\| |\psi\rangle\right\|\neq 0$), then 
        \begin{equation}
                0=\langle\psi|\sum_{l=1}^M
                \lambda_l\hat{J}_l^{\dagger}\hat{J}_l\,|\psi\rangle= \sum_{l=1}^M\lambda_l
                \left\|\hat{J}_l |\psi\rangle \right\|^2.
        \end{equation}
        Because $\lambda_l>0$ we can conclude that $\hat{J}_l
        |\psi\rangle=0|\psi\rangle$ $\forall $
        $l$. Proposition~\ref{prop:DFS0} then guarantees that
        $|\psi\rangle$ is  in an IPDFS 
\end{proof}

\begin{remark}
        As the eigenvalues of the decoherence operator $\hat{\Gamma}$ are the inverse lifetimes of
        the corresponding eigenstates, Proposition~\ref{prop:DFS} shows that a state
        $|\psi\rangle$ is in an IPDFS iff it has an infinite lifetime.
\end{remark}

One purpose of this work is to study whether or not an IPDFS (and consequently DFS) is possible for particles located at
different positions. We therefore have to distinguish IPDFS (DFS), 
which require two or more particles from those that already exist for
a single particle, i.e. from internal IPDFS (DFS) states.
\begin{definition}
       The single-particle IPDFS ${\cal H}_{\text{s-IPDFS}}^{(n)} \subset {\cal H}_n$
        of particle $n$ is spanned by all states which are
        decoherence-free if the system consists only of particle $n$, i.e., if
        the sum in the Hamiltonian (\ref{hInt}) contains only one particle.
        The single-particle IPDFS for the complete $N$-particle system is the tensor
        product of the individual single-particle IPDFS,
        $\mathcal{H}_{\text{s-IPDFS}}\equiv\otimes_{n=1}^N \mathcal{H}_\text{IPDFS}^{(n)}$.
\end{definition}
\begin{definition}
        The multi-particle DFS is the complement
        $\mathcal{H}_{\text{m-IPDFS}}=\mathcal{H}_{\text{IPDFS}}\setminus \mathcal{H}_{\text{s-IPDFS}}$.
\label{def:sipdfs}
\end{definition}
The single-particle and multi-particle DFS can be defined similarly. 
The single-particle IPDFS (DFS) corresponds to a tensor product of IPDFSs (DFSs) for each
particle. For instance, if all particles are atoms prepared in their ground states the corresponding $N$-particle state is stable against decoherence due to spontaneous decay.
The multi-particle IPDFS (DFS) typically involves states in which some or all particles are entangled and relies on destructive interference of the transition amplitudes for energy transfer from one particle to another. In the next section we introduce reasonable assumptions about realistic reservoirs to study whether or not a multi-particle IPDFS (DFS) can exist.

\section{IPDFS in realistic systems}
\label{sec:DFS}

\subsection{Markov-Born master equation}
\label{subsec:Markov-Born}

In a realistic system, the locations and spatial configuration of the particles are important.
No two particles can
occupy the same point, i.e.~$\mathbf{x}_n\neq\mathbf{x}_{n'}$ for $n\neq n'$. However,
particles may be very close together and the limit that 
$\mathbf{x}_n=\mathbf{x}_{n'}$ $\forall$ $n,n'$ is referred to as the Dicke limit~\cite{Dic54}.
The reservoir operators $\hat{E}^i(\mathbf{x})$ introduced above are position-dependent,
and the reservoir Hamiltonian~$\hat{H}_\text{R}$ governs the dynamics of these operators.
For example, the set  $\{\hat{E}^i\}$ corresponds to the vector components of
the electric and magnetic fields for the case of an electromagnetic
reservoir.

The system+reservoir interaction is given by Eq.~(\ref{hInt}),
which yields the master equation in the Markov-Born approximation:
\begin{align} 
        \dot{\rho}(t)=&-i[\hat{H}_{S}, \rho]
        -\int^{\infty}_0 \text{d}\tau\sum_{n,m,i,j}
        \Big( \left[ \hat{d}_n^i,\hat{d}_m^j(-\tau)\, \rho\right]  \nonumber \\ &\times
        \left\langle \hat{E}^i(\mathbf{x}_n,\tau)\, 
        \hat{E}^j(\mathbf{x}_m, 0)\right\rangle+ \left[\rho \, \hat{d}_m^j(-\tau)\,   
        \hat{d}_n^i\right] \nonumber \\ & \times
        \left\langle \hat{E}^j(\mathbf{x}_m,0)\hat{E}^i(\mathbf{x}_n,\tau)\right\rangle\Big).
\label{masterEq}
\end{align} 
In this master equation
\begin{align}
  \hat{d}_n^i(-\tau)=&\exp(-i\hat{H}_\text{S}\tau)\hat{d}_n^i\exp(i\hat{H}_\text{S}\tau)\nonumber \\
  =&\sum_{\stackrel{\alpha, \beta;}{\omega_{\alpha\beta}\neq 0}}\hat{\sigma}_{n,\alpha\beta}
  \text{e}^{-i\tau\omega_{\alpha\beta}} \, d^i_{\alpha\beta},
\label{eq:dntau}
\end{align}
and  $ \hat{E}(\tau)=U_\text{R}^\dagger \hat{E} U_\text{R}$ 
are the operators in the interaction picture,
with $U_\text{R}=\exp( -i\hat{H}_\text{R}\tau)$.

\subsection{Properties of the reservoir}
\label{subsec:reservoir}
Master equation~(\ref{masterEq}) describes the evolution of $N$ particles interacting with a
reservoir, whose propagation can be described by the correlation function 
\begin{align}
        G^{ij}_{nm}(\tau) 
                = &\theta(\tau)\left\langle\hat{E}^i(\mathbf{x}_n,\tau)\hat{E}^j(\mathbf{x}_m,0)\right\rangle
                        \nonumber       \\
                = & \theta(\tau) \text{Tr}_{R}\left ( \rho_{R}\, \hat{E}^i(\mathbf{x}_n,\tau)
                        \hat{E}^j(\mathbf{x}_m,0) \right),
\label{Eq:cf}
\end{align}
with  $\rho_\text{R}$ the density operator for the
initial state of the reservoir. In what follows we assume that
the reservoir has the following reasonable properties. The last  two points
below are standard assumptions 
in open system theory using Markovian master equations, which we state here explicitly for completeness.
\begin{enumerate}[(i)]
\item   $\hat{H}_\text{R}$ is time-independent.
\item   The reservoir is initially in a stationary state:
  $[\rho_{R}, \hat{H}_\text{R}]=0$.
  Then~$\rho_\text{R}$ is diagonal in the
  $\hat{H}_\text{R}$-eigenbasis~$\{|\psi_{r \mathbf{k}}\rangle\}$, 
  with $r$ and $\mathbf{k}$ quantum numbers for these states
  (the meaning of ${\bf k}$ is explained in assumption (iii))
  and $\{\Omega_{r\mathbf{k}}\}$ the 
  corresponding eigenvalues:
  \begin{equation} 
    \rho_\text{R} = \int \text{d}^3\mathbf{k}\; \sum_\text{R}  p_{r \mathbf{k}} 
    |\psi_{r \mathbf{k}}\rangle \langle \psi_{r \mathbf{k}}|,
  \end{equation} 
  with $\{p_{r \mathbf{k}}\}$ the probabilistic weighting of these eigenstates.
\item   The reservoir is translationally invariant: for~$\hat{\mathbf{P}}$ the
  generator of translations for the reservoir in three dimensions,  
  $[\hat{H}_\text{R} , \hat{\mathbf{P}} ] = [\rho_\text{R} , \hat{\mathbf{P}} ] =0$, and the
  spectrum of $ \hat{\mathbf{P}}$ is continuous.
  Typically $\hat{\mathbf{P}} $ is the total momentum operator of the reservoir.
  The vector quantum number~$\mathbf{k}$ thus has meaning as the momentum:
  $\hat{\mathbf{P}}|\psi_{r \mathbf{k}}\rangle=\mathbf{k}|\psi_{r
    \mathbf{k}}\rangle$.
\item   $\hat{H}_\text{R}$ is a function of  
  $\left|\hat{\mathbf{P}}\right|$ and independent of the orientation of $\hat{\mathbf{P}}$.
  This assumption is widely valid. For example a
  reservoir of~$M$ interacting particles with Hamiltonian
  \begin{equation}
    \hat{H}_\text{R} = 
    \sum_i \frac{\hat{\mathbf{p}}_i^2}{2m} + \sum_{j>i} V(\mathbf{x}_i-\mathbf{x}_j)
  \end{equation}
  has a corresponding generator of translations 
  $\hat{\mathbf{P}} = \sum_i\mathbf{p}_i$.
  Expressed in relative and center-of-mass coordinates, the Hamiltonian becomes
  \begin{equation}
    \hat{H}_\text{R} =\frac{\hat{\mathbf{P}}^2}{2 M m} + \hat{H}_\text{rel},
  \end{equation}
  with $\hat{H}_\text{rel}$ depending only on the relative coordinates and not
  on the orientation of $\hat{\mathbf{P}}$. 
\item   Translations of the reservoir operators $\hat{E^i}(\mathbf{x})$ 
are generated by the operator $\hat{\mathbf{P}}$,
\begin{equation} 
   \hat{E}^i(\mathbf{x}_0 +\mathbf{x} ) =
   \text{e}^{i\hat{\mathbf{P}}\cdot\mathbf{x}}\hat{E}^i(\mathbf{x}_0)  
   \text{e}^{-i\hat{\mathbf{P}}\cdot \mathbf{x}} \; .
\end{equation} 
\item   For technical reasons we need  to introduce the set of $\mathbf{k}$ vectors
\begin{equation}
   K_{\omega,r,s}(\mathbf{k}')=\left\lbrace
         |\mathbf{k}|:\, 
   \Omega_{r\mathbf{k}} =\Omega_{s\mathbf{k}'} + \omega)\right\rbrace,
\label{eq:set K} 
\end{equation} and require that the reservoir transformation function
\begin{align}
\label{eq:polyk}
        \left\langle\psi_{s \mathbf{k}'}\left|\hat{E^i}(0)\right|\psi_{r \mathbf{k}}\right\rangle 
\left \langle \psi_{r \mathbf{k}}\left|\hat{E^j}(0)\right|\psi_{s \mathbf{k}'}\right\rangle
         \Big|_{|\mathbf{k}|\in K_{\omega,r,s}(\mathbf{k}')}
\end{align} 
 is either (a)~a polynomial in terms of Cartesian components of $\mathbf{k}$, or (b)~is equal to $f(\mathbf{k})A^{ij}_{rs}$ such that  $A^{ij}_{rs}$ independent of
$\mathbf{k}$ and the support of $f(\mathbf{k})$ contains $|\mathbf{k}|=r$ for some $r\in K_{\omega,r,s}(\mathbf{k}')$. The transformation function describes the change in the state of the reservoir when a transition in the system occurs that changes its energy by $\hbar \omega$. The expression in Eq.~(\ref{eq:polyk}) is constrained to the set $K_{\omega,r,s}(\mathbf{k}')$ because only transitions that conserve energy in the overall system (quantum system+reservoir) are important. Case (b) describes a reservoir composed from scalar particles. 
Note that there exists $k\in K_{\omega,r,s}(\mathbf{k}')$ for some $\omega, r,s,\mathbf{k}'$ such that the transformation function~(\ref{eq:polyk}) is not identically equal to zero for $|\mathbf{k}|=k$, since otherwise the quantum system would not be coupled to the reservoir.
\item   All energy-nonconserving transitions in the system of $N$ identical
  particles average to zero on the time scale relevant to the Markovian master
  equation. This condition guarantees that the Markovian master equation we
  derive is completely positive (i.e. physical) \cite{Dum79}. Inclusion of
  energy-nonconserving transitions modifies $\hat{\Delta}$ \cite{Agar71,
    Knight, Agar73}, which is irrelevant for the purpose of this paper since,
  as we have shown, only the decoherence operator $\hat{\Gamma}$ is needed for
  determining IPDFS.
\item   System+reservoir coupling is sufficiently weak so that
second-order perturbation theory is valid.
\item   The reservoir correlation function decays rapidly in
time so that backreaction is effectively instantaneous, which underpins
the Markovian master equation and guarantees that the frequency spectrum of $G$ contains
only frequencies that are much larger than the coupling parameters.
Consistency of the master equation then implies that the processes
underlying it are energy conserving~\cite{Dum79}.
\end{enumerate}

\begin{definition}
A homogeneous isotropic Markovian reservoir is any reservoir that satisfies all conditions~(i) to~(ix) above.
\end{definition}

From these assumptions, the Fourier transform of the correlation function~$G$ is
\begin{align}
\label{eq:Gnm}
        \widetilde{G}_{nm}^{ij}(\omega)=&   
        \sum_{r,s}\int \text{d}^3\mathbf{k}' \; \int \text{d}^3\mathbf{k} \; p_{s \mathbf{k}'}\,
        \text{e}^{i(\mathbf{k}-\mathbf{k}')\cdot(\mathbf{x}_n-\mathbf{x}_m)} \nonumber \\&\times 
        \left\langle\psi_{s \mathbf{k}'}\left|\hat{E^i}(0)\right|\psi_{r \mathbf{k}}\right\rangle \; 
        \left\langle \psi_{r \mathbf{k}}\left|\hat{E^j}(0)\right| \psi_{s \mathbf{k}'}\right\rangle 
        \nonumber\\ &\times  \left(\pi \delta(\omega -\Omega_{r \mathbf{k} s \mathbf{k}'}) 
        +i\frac{\cal P}{\omega-\Omega_{r \mathbf{k} s \mathbf{k}'}}\right).
\end{align}
Here
\begin{equation}
        \Omega_{r \mathbf{k} s \mathbf{k}'}=\Omega_{r \mathbf{k}}-\Omega_{s \mathbf{k}'}
\end{equation}
and details of the calculation are provided in Appendix~\ref{app:Fourier}.

\subsection{Decoherence operator and correlation function}
After comparing Eqs.~(\ref{Eq:generalME}) to~(\ref{masterEq}) and recalling  the definition for the correlation  function $G_{nm}^{ij}(\tau)$ from Eq.~(\ref{Eq:cf}), one can identify the non-Hermitian transition operator specific to our system as 
\begin{align}
\label{eq:A}
i\hat{\Delta} +\frac{1}{2}\hat{\Gamma}=\int_{-\infty}^\infty \text{d}\tau
  \sum_{\stackrel{n,m;}{i,j}}G_{nm}^{ij}(\tau) \, \hat{d}_n^i \, \hat{d}_m^j(-\tau).
\end{align}
From Eq.~(\ref{eq:dntau}), we can rewrite Eq.~(\ref{eq:A}) as
\begin{align}
\label{eq:Gamma} 
        i\hat{\Delta} +\frac{1}{2}\hat{\Gamma}
        =& \sum_{ \stackrel{n,m;}{ i,j }} \sum_{\stackrel{\alpha,\beta;}
        {\omega_{\alpha\beta}\neq 0}}\sum_{\stackrel{\beta',\alpha';}
        {\omega_{\alpha'\beta'}\neq 0}}  d^i_{\alpha\beta}\, d^j_{\beta'\alpha'} \, 
        \hat{\sigma}_{n,\alpha\beta} \, \hat{\sigma}_{m,\beta'\alpha'}\nonumber \\&\times
        \int_{-\infty}^\infty \text{d}\tau \, G_{nm}^{ij}(\tau) \text{e}^{-i\tau\omega_{\beta'\alpha'}}.
\end{align}
By comparing the integral in Eq.~(\ref{eq:Gamma}) to the integral in Eq.~(\ref{Eq:ft}), one sees that 
\begin{align}
        \int_{-\infty}^\infty \text{d}\tau \, G_{nm}^{ij}(\tau) 
        \text{e}^{-i\tau\omega_{\beta'\alpha'}}= \widetilde{G}_{nm}^{ij}(-\omega_{\beta'\alpha'}).
\end{align}

Now we introduce the collective operator
\begin{align}
        \hat{\Sigma}_{rs}\left(\mathbf{k}, \mathbf{k}', \omega\right) = \sum_n
        e^{i(\mathbf{k}-\mathbf{k}')\cdot
        \mathbf{x}_n}  \hat{\Sigma}_{rs;\,n}\left(\mathbf{k}, \mathbf{k}', \omega\right),
\label{eq:Sigma}
\end{align}
with  
\begin{align} 
   \hat{\Sigma}_{rs;n}\left(\mathbf{k}, \mathbf{k}', \omega\right)=&
   \sum_{\stackrel{{\scriptstyle \alpha, \beta}}{\omega_{\alpha\beta}=\omega}}
    \hat{\sigma}_{n,\alpha\beta} \sqrt{p_{s \mathbf{k}'}}\nonumber \\&\times 
   \left\langle\psi_{r \mathbf{k}}\left|\sum_i\hat{E^i}(0)d^i_{\alpha\beta} \,
   \right|\psi_{s \mathbf{k}'}\right\rangle.
\end{align} 
Here $\hat{\Sigma}_{rs;\,n}\left(\mathbf{k}, \mathbf{k}', \omega\right)$ is an
operator that excites ($\omega >0$) or de-excites ($\omega<0$)
the $n^\text{th}$ particle. 

Operator~(\ref{eq:Sigma}) is a direct sum of operators on each particle and
incorporates the reservoir state and coupling term to create a specific
raising or lowering operator. The operator is defined on states
$|\alpha\rangle$ and $|\beta\rangle$ connected by raising (or, if $\omega<0$,
lowering) operators $\hat{\sigma}_{n,\alpha\beta}$ so that the energy
difference between these states is $\omega$. Somewhat analogous to the number
operator, we combine $\Sigma$ with its adjoint~$\Sigma^\dagger$ to define
\begin{equation}
\label{eq:Upsilon_n}
        \hat{\Upsilon}_{rs;\,n}\left(\mathbf{k}, \mathbf{k}', \omega\right)
                \equiv  \hat{\Sigma}_{rs;\,n}^{\dagger}\left(\mathbf{k}, \mathbf{k}', \omega\right)
                 \hat{\Sigma}_{rs;\,n}\left(\mathbf{k}, \mathbf{k}', \omega\right)
\end{equation}
for the $n^\text{th}$ particle, and
\begin{equation}
\label{eq:Upsilon}
        \hat{\Upsilon}_{rs}\left(\mathbf{k}, \mathbf{k}', \omega\right)
                \equiv   \hat{\Sigma}_{rs}^{\dagger}\left(\mathbf{k}, \mathbf{k}', \omega\right)
                 \hat{\Sigma}_{rs}\left(\mathbf{k}, \mathbf{k}', \omega\right)
\end{equation}
globally.

Henceforth, we impose assumption (vii) of Sec.~\ref{subsec:reservoir}, which
in this context  requires that all terms in Eq.~(\ref{eq:Gamma}) that fail
condition $\omega_{\alpha\beta}=\omega_{\alpha'\beta'}$ can be neglected. This
assumption ensures the consistency of the Markovian master
equation~\cite{Dum79}. 
Then by combining Eq.~(\ref{eq:Gnm}) with Eq.~(\ref{eq:Gamma}) and retaining
only energy-conserving terms, we obtain 
\begin{align} 
        i\hat{\Delta}+\frac{1}{2}\hat{\Gamma}= \sum_{r,s}\int \text{d}^3\mathbf{k}' \; \int \text{d}^3\mathbf{k}
         \sum_{\omega_{\alpha\beta}\neq 0} 
         \hat{\Upsilon}_{rs}(\mathbf{k}, \mathbf{k}', \omega_{\alpha\beta})
                \nonumber \\ \times
         \left( \pi \delta(\omega_{\alpha\beta} -\Omega_{r \mathbf{k} s \mathbf{k}'})
         +i\frac{\cal P}{\omega_{\alpha\beta}-\Omega_{r \mathbf{k} s \mathbf{k}'}}\right).
\label{eq:gammaExtract}
\end{align} 

The decoherence operator $\hat{\Gamma}$ is the  Hermitian part of the  right-hand side of 
Eq.~(\ref{eq:gammaExtract}). Because operator $\hat{\Upsilon}_{rs}(\mathbf{k},
\mathbf{k}', \omega_{\alpha\beta})$ is Hermitian by construction, it follows
that the term involving the Dirac~$\delta$ generates the Hermitian decoherence
operator, whereas the term involving the principal value ${\cal P}$ 
generates the energy level shifts (the ``Lamb shift'' in quantum optics). We thus find
\begin{align}
        \hat{\Gamma}_\omega =& 2\pi \sum_{r,s}\int \text{d}^3\mathbf{k}' \; \int \text{d}^3\mathbf{k}
         \hat{\Upsilon}_{rs}(\mathbf{k}, \mathbf{k}', \omega)
                \nonumber \\ & \times
         \delta(\omega -\Omega_{r \mathbf{k} s \mathbf{k}'}).
\end{align}
and
\begin{equation}
        \hat{\Gamma}=\sum_{\omega_{\alpha\beta}\neq 0} \hat{\Gamma}_{\omega_{\alpha\beta}}
\end{equation}
By assumption (iv) of Sec.~\ref{subsec:reservoir}, the frequency difference
$\Omega_{r \mathbf{k} s \mathbf{k}'}$ depends on the absolute value of 
$\mathbf{k}$ only. Therefore the Dirac~$\delta$  
fixes the absolute value of $|\mathbf{k}|$ by the condition
\begin{equation}
\label{fix k}
        \Omega_{r \mathbf{k} s \mathbf{k}'} = \omega_{\alpha'\beta'} =
        \omega_{\alpha\beta};
\end{equation} 
that is, the energy differences in the reservoir
$\Omega_{r\mathbf{k}}-\Omega_{s\mathbf{k}'}$ equal the corresponding energy differences for
the system transition $\omega_\alpha-\omega_\beta$. For a given vector
$\mathbf{k}'$ and energy difference $\omega_{\alpha\beta}$ Eq.~(\ref{fix k}) 
determines a set of values  $|\mathbf{k}|$ for which it is fulfilled.
We denote this set by 
\begin{equation}
   K_{\omega,r,s}(\mathbf{k}')=\left\lbrace
         |\mathbf{k}|:\, 
   \Omega_{r\mathbf{k}} =\Omega_{s\mathbf{k}'} + \omega)\right\rbrace,
\label{eq:set Ka} 
\end{equation}
which is not empty for some $\omega$; otherwise the system and reservoir would
never exchange energy. 

Let ${\cal S}_k$ denote a sphere of radius $k$. Then the integral with respect
to $\mathbf{k}$ in the expression for the decoherence operator $\hat{\Gamma}$
is reduced to an integral over the surface of the sphere ${\cal S}_k$ with
$k\in K_{\omega,r,s}(\mathbf{k}')$, where $\omega$ takes  all possible frequency
differences within one particle: 
\begin{equation}
        \hat{\Gamma}=2\pi \sum_{\stackrel{r,s}{\omega_\neq 0}}
        \int \text{d}^3\mathbf{k}'\sum_{k\in K_{\omega,r,s}(\mathbf{k}')} 
        \int_{ \mathbf{k}\in{\cal S}_k} 
        \text{d}^2\mathbf{k}\; \hat{\Upsilon}_{rs}\left(\mathbf{k}, \mathbf{k}', \omega\right).
\label{eq:gamma}
\end{equation}

Although the domain for quantum numbers $r$, $s$, and $\mathbf{k'}$ could be
discrete, 
continuous, or a combination thereof, assumptions \emph{(i-ix)} imply that $\mathbf{k}\in{\cal S}_k$,
which is a sphere of radius~$k$. We have thereby shown that decoherence operator
$\hat{\Gamma}$ is related to $\hat{\Upsilon}$ and determined the
general expression for $\hat{\Sigma}$. This information will allow us to
demonstrate in the next section that multi-particle IPDFS (and consequently multi-particle DFS) cannot exist in this
system.

\section{Conditions for multi-particle DFS states}
We seek a necessary criterion for the existence of multi-particle DFS
states. We do this by establishing necessary and sufficient conditions for existence of multi-particle IPDFS for spatially separated particles in a homogeneous isotropic Markovian reservoir. We  start with excluding the single-particle IPDFSs and we prove the following Proposition.

\begin{Proposition}
\label{prop:noMDFS}
   There is no multi-particle DFS for spatially separated particles in a homogeneous isotropic Markovian reservoir.
\end{Proposition} 
 
\begin{proof}
In this proof we make use of Proposition \ref{prop:DFS}. Therefore we need to
check whether our system satisfies the hypothesis of Proposition
\ref{prop:DFS}, i.e. whether all eigenvalues of $\hat{J}_l$ are zero. Each
operator $\hat{\Gamma}_\omega$ can be decomposed as in
Eq.~(\ref{eq:decoherencematrix}). Index $l=(n,\,\omega, \alpha,\,\beta)$ for
operator $\hat{J}_l$ depends on $n$ (particle index), on  $\omega$ (possible
frequency differences within one particle), and on energy levels labeled by $\alpha$ and $\beta$ such
that $\omega_{\alpha\beta}$ is constrained to equal $\omega$. Each operator
$\hat{J}_l$ is some linear combination of operators
$\hat{\sigma}_{n,\alpha\beta}$ with $\alpha$ and $\beta$ constrained so that
$\omega_{\alpha\beta}$ is some fixed non-zero value. This means that there
exists a representation with all operators $\hat{J}_l$ either upper or lower
triangular matrices with zeros along the diagonal. Then we know that all
eigenvalues of operators $\hat{J}_l$ are equal to zero and Proposition
\ref{prop:DFS} applies to our system.  
 
First we need to determine the condition for existence of single-particle IPDFS
states and remove these from the subspace
$\mathcal{H}_{\text{IPDFS}}$. According to the definition~\ref{def:sipdfs} and Proposition
\ref{prop:DFS}, we are seeking states $|\alpha\rangle\; \in \; \mathcal{H}_n$
such that $\left\|\hat{\Gamma}_n|\alpha\rangle\right\|=0$, with
$\hat{\Gamma}_n$ the decoherence operator $\hat{\Gamma}$ restricted to the
$n^{\text{th}}$ particle. At the same time, we know that 
\begin{equation}
        \hat{\Gamma}_n=2\pi \sum_{\stackrel{r,s}{\omega_\neq 0}}
                \int \text{d}^3\mathbf{k}'\sum_{k\in K_{\omega,r,s}(\mathbf{k}')} 
        \int_{ \mathbf{k}\in{\cal S}_k} 
        \text{d}^2\mathbf{k} 
        \hat{\Upsilon}_{rs;\,n}\left(\mathbf{k}, \mathbf{k}', \omega\right)\nonumber,
\end{equation}
As a result,  a state 
$|\alpha\rangle\;\in\; \mathcal{H}_{\text {s-IPDFS}}^{(n)}=\mathcal{H}_n\cap\mathcal{H}_\text{IPDFS}$ iff
\begin{equation}
   \left\|\hat{\Sigma}_{rs;\,n}\left(\mathbf{k}, \mathbf{k}', \omega\right)|\alpha\rangle\right\|=0
\end{equation}
for all the external quantum numbers $r$, $s$, $\mathbf{k}'$, for all possible
frequency differences $\omega$ within one particle, for all values $k$ in the
set $K_{\omega,r,s}(\mathbf{k}')$, and for all vectors $\mathbf{k}$ that lie
in the sphere ${\cal S}_k$. The space of single-particle IPDFS of the total
system is then ${\cal H}_\text{s-IPDFS} = \oplus_{n=1}^N \mathcal{H}_{\text {s-IPDFS}}^{(n)}$

Now we determine under what conditions the multi-particle IPDFS
$\mathcal{H}_{\text {m-IPDFS}}=\mathcal{H}_{\text{IPDFS}}\setminus\mathcal{H}_{\text{s-IPDFS}}$
is not empty. By an analogous argument we infer that 
\begin{equation}
        |\alpha\rangle\,\in\,\mathcal{H}_{\text{IPDFS}}\,\,\Rightarrow\,\,
        \left\|\hat{\Sigma}_{rs}\left(\mathbf{k}, \mathbf{k}', \omega\right)|\alpha\rangle\right\|=0
\end{equation}
for all quantum numbers $r$, $s$, $\omega$,  $\mathbf{k}'$, $k$ and for all
vectors $\mathbf{k}$ that are in the sphere ${\cal S}_k$. 

Let $|y_1\rangle, \ldots, |y_{N'}\rangle$ be a complete basis for $\mathcal{H}_{\text{m-IPDFS}}$. Then
\begin{align}
\label{spDFS}
  \sum_n &e^{i(\mathbf{k}-\mathbf{k}')\cdot 
   \mathbf{x}_n}  \hat{\Sigma}_{rs;\,n}
   \left(\mathbf{k}, \mathbf{k}', \omega\right)|y_j\rangle = 0 \; 
\nonumber \\&\forall\;j=1, \ldots,N' \; ,
\end{align}
with $N' = \text{dim}(\mathcal{H}_{\text{m-IPDFS}})$. By assumption
not all $\hat{\Sigma}_{rs;\,n}$ do annihilate the state $|y_j\rangle$ because it
then would be an element of $\mathcal{H}_{\text{s-IPDFS}}$. 

Eq.~(\ref{spDFS}) implies in particular that 
\begin{align}
\label{mpDFS}
        &\sum_n \text{e}^{i(\mathbf{k}-\mathbf{k}')\cdot( 
        \mathbf{x}_n-\mathbf{x}_m)} \nonumber \\
        \times&  \left\langle y_j\right| \hat{\Sigma}^\dagger_{rs;\,m}(\mathbf{k}, \mathbf{k}', 
        \omega)\hat{\Sigma}_{rs;\,n}\left(\mathbf{k}, \mathbf{k}', \omega\right)\left|y_j\right\rangle\Big|_{|\mathbf{k}|\in K_{\omega,r,s}(\mathbf{k}')}
        =0 
\end{align}
for all  $j=1,\ldots,N'$, for all  quantum numbers $r$, $s$, $\omega$,
$\mathbf{k}'$, $k$, for all vectors $\mathbf{k}$ that are in the sphere 
${\cal S}_k$, and for all $m=1,\ldots, N$. Note that Eq.~(\ref{mpDFS}) is the kernel $\hat{\Upsilon}_{rs}\left(\mathbf{k}, \mathbf{k}', \omega\right)$ needed to calculate operator $\hat{\Gamma}$ as in Eq.~(\ref{eq:gamma}).
Condition~(\ref{mpDFS}) is not trivially satisfied because for each $j$ 
there is at least one $m$ such that 
\begin{equation}
        \left\langle y_j\left| \hat{\Upsilon}_{rs;\,m}\left(\mathbf{k}, \mathbf{k}', \omega\right)
        \right|y_j\right\rangle =\left\|\hat{\Sigma}_{rs;\,m}\left(\mathbf{k},
          \mathbf{k}', \omega\right)|y_j\rangle\right\|^2\neq 0, 
\end{equation}
since $|y_j\rangle\,\notin\,\mathcal{H}_{\text{s-IPDFS}}$.
Let
\begin{align}
        &c_{m,j}^n = \nonumber\\ &\left\langle y_j\left|\hat{\Sigma}^\dagger_{rs;\,m}\left(\mathbf{k}, \mathbf{k}', \omega\right)
           \hat{\Sigma}_{rs;\,n}\left(\mathbf{k}, \mathbf{k}', \omega\right)\right|y_j\right\rangle\Big|_{|\mathbf{k}|\in
         K_{\omega,r,s}(\mathbf{k}')}.
\end{align}
We note that there exists $k\in K_{\omega,r,s}(\mathbf{k}')$ such that $c_{m,j}^n\neq 0$ since otherwise quantum system would not be coupled to the reservoir. 
Then $\mathcal{H}_{\text {m-IPDFS}}$ is non-empty only if
\begin{equation}
\label{mpDFS2}
        \sum_n \text{e}^{i\mathbf{k}\cdot\mathbf{x}_n}  c_{m,j}^n=0
\end{equation}
is satisfied for all possible  vectors~$\mathbf{k}$ that are in the sphere ${\cal S}_k$, for all $j=1,\ldots,N'$,  and for all $m=1,\ldots,N$ with not all $c_{m,j}^n=0$. Now we make use of the two alternative assumptions~(vi, a) and (vi, b) in Subsec.~\ref{subsec:reservoir}.

\emph{Assumption~(vi, a):--} This assumption allows us to conclude that
\begin{equation}
        c_{m,j}^n=h^n_{m,j}(\mathbf{k})b^n_{m,j},
\end{equation}
with $h^n_{m,j}(\mathbf{k})$ a non-zero polynomial function in components of vector $\mathbf{k}$  defined on the sphere ${\cal S}_k$ and $b_{m,j}^n$ coefficients independent of $\mathbf{k}$.
Then Eq.~(\ref{mpDFS2}) becomes
\begin{equation}
        \sum_n \;h^n_{m,j}(\mathbf{k})\text{e}^{i\mathbf{k}\cdot\mathbf{x}_n}  b_{m,j}^n=0,\,\, \mathbf{k}\in{\cal S}_k,
\end{equation} 
which is equivalent to determining conditions when a set of functions 
\begin{equation}
        \{h^n_{m,j}(\mathbf{k})\text{e}^{i\mathbf{k}\cdot\mathbf{x}_n}\}_{n=1}^N
\end{equation}
defined on a sphere ${\cal S}_k$ can be made linearly dependent.  It is well
known that functions of this form  are linearly independent unless there
exists $m\neq n$ such that $\mathbf{x}_n=\mathbf{x}_m$ (see Appendix
~\ref{app:liIn}). Thus requirement (\ref {mpDFS}) can only be fulfilled if
$\exists \;m\neq n$ such that $\mathbf{x}_n=\mathbf{x}_m$, i.e. at least two
particles occupy the same point in space. 
 
\emph{Assumption~(vi, b):--} This assumption yields
\begin{equation}
        c_{m,j}^n=f(\mathbf{k})b^n_{m,j},
\end{equation}
 and $\{b_{m,j}^n\}$ are coefficients independent of $\mathbf{k}$.

As the function $f(\mathbf{k})$ is not zero for $\mathbf{k}\in {\cal S}_k$, condition (\ref{mpDFS2}) is equivalent to requiring that 
\begin{equation}
\label{mpDFS3}
        \sum_n \text{e}^{i\mathbf{k}\cdot\mathbf{x}_n}  b_{m,j}^n=0
\end{equation}
for all possible  vectors $\mathbf{k}\in{\cal S}_k$ with not all $b_{m,j}^n=0$. 
This is a special case of the situation we considered above and thus we reach the same conclusion as before: a multi-particle IPDFS can exist  only when  at least two particles occupy the same point in space. Consequently, a multi-particle DFS also does not exist when no two particles occupy the same point in space.
\end{proof}

Although Proposition~\ref{prop:noMDFS} rules out a multi-particle DFS, states
that do not decohere over a long lifetime, i.e.  significantly enhanced beyond
the longest lifetimes of the individual particles, may be just as
practical. Because of the nature of the reservoir being continously
translationally invariant (assumption (iii)), we expect that this significant enhancement is
achieved when particles are close together on some scale determined by the
reservoir; we explore this case in the next section.

\section{Decoherence suppression}
\label{sec:approximate}

We now apply the formalism developed in previous sections to a well-studied model in quantum optics, namely a set of two-level atoms interacting with a multi-mode radiation field,
which serves as a basis for studying collective phenomena such as superradiance and subradiance \cite{Dic54, Rehler, Bonifacio, Gross}.
In fact some of the notation was inspired by this example, such as using the variable $d$, which
is typically used for the electric dipole moment of the atoms, and $E^i$,
which is used for components of the electric field. 
In general the reservoir for a collection of atoms, such as  a gas, closely
satisfies the homogeneous isotropic Markovian conditions, and the
system+reservoir coupling is as described in previous sections. 

This model also provides a welcome simplicity in that each atom contains only
one excited state so an internal DFS is not possible; hence there is no
non-trivial single-particle DFS (i.e. other than the ground state for the system).
As atoms cannot be truly co-located, we know from Proposition~\ref{prop:noMDFS}
that in this system a perfect DFS exists only in the Dicke limit for which atoms are co-located.

The question we address now is whether an approximate DFS can exist, and how well it behaves,
in conditions that are close to the ideal condition required for the DFS to exist. The atomic case gives us a 
concrete and well-studied example that gives insight to answer this question.

Let us begin by assuming that the two-level atoms are located at positions
$\mathbf{x}_j$ ($j=1, \ldots, N$)  
and that atomic dipole moments are given by $\mathbf{d}$. 
Then the decoherence operator has the form \cite{Bel69, Lehm1, Lehm2, Agar70, agar74} 
\begin {equation}
        \hat{\Gamma}=\sum_{j,k=1}^N \gamma_0\gamma_{jk} \hat{\sigma}_j^\dagger \hat{\sigma}_k
\end{equation}
with
\begin{align}
\gamma_ {jk}\equiv & \frac{3}{2}\Bigg\{\left[ 1-\left(\frac{\mathbf{d}\cdot
      \mathbf{x}_{jk}}{|\mathbf{d}|\,|\mathbf{x}_{jk}|}\right)^2\right]
\frac{\sin k_0 x_{jk}}{k_0 x_ {jk}} 
\nonumber \\ 
   &+\left[ 1-3\left(\frac{\mathbf{d}\cdot
      \mathbf{x}_{jk}}{|\mathbf{d}|\,|\mathbf{x}_{jk}|}\right)^2\right] \left(
  \frac{\cos k_0 x_{jk}}{(k_0 x_{jk})^2}-\frac{\sin k_0 x_{jk}}{(k_0
    x_{jk})^3}\right)\Bigg\}, 
\end{align}
and $k_0=2\pi/\lambda_0$, ${\bf{x}}_{jk}={\bf{x}}_j-{\bf{x}}_k$, $x_{jk}=|{\bf{x}}_{jk}|$, 
$\hat{\sigma}_k$ is the lowering operator for the $k^{\text{th}}$ atom,
$\gamma_0$ is the Einstein $A$ coefficient,  and $\lambda_0$ is the resonant
wavelength. We also define reduced decoherence matrix to be
$\overleftrightarrow{\gamma}=(\gamma_{jk})$. 

Eigenvalues of the decoherence operator, $\hat{\Gamma}$, provide inverse
lifetimes of the states in the system. For many applications, it is sufficient
to look at the reduced matrix $\overleftrightarrow{\gamma}$. This matrix describes the
states in the system of $N$ two-level atoms with one atom in the excited
state. It also describes the states for $N$ two-level atoms with all atoms
except one in the excited state (see Appendix \ref{app:reducedmatrix}). In
general, it  is much easier to deal with an $N\times N$ matrix
$\overleftrightarrow{\gamma}$ rather than with an $2^N\times 2^N$ matrix
representation of the decoherence operator $\hat{\Gamma}$. 

For special cases where the location of atoms can be described with one variable,
one can produce two dimensional plots that show the dependence of lifetimes on
the separation between atoms. As an example, consider $N$ atoms in a line. The
separation between adjacent atoms is taken to be equal ($x_{i,i+1}=r$). Figs.~1(a, c, d) show  particular configurations of two-level atoms in a line as well as the orientation of the dipole moment for the atoms that we consider here.  Figs.~2(a, c, d) depict the inverse lifetimes  (eigenvalues
of matrix $\overleftrightarrow{\gamma}$) of the states in the system of two
and four atoms, when only one atom is excited, as a function of $r$. In
Fig.~1(b) we consider the system of four atoms that form a square.  The
inverse lifetimes for states that arise when only one atom is excited in this
system are shown in Fig.~2(b). From the plots one can see that there are
long-lived states (i.e., inverse lifetimes close to zero) when adjacent atoms
are separated by less than a quarter of the emission wavelength
($\lambda_0/4$). For each cluster of closely located atoms, one could use the
states with  longest lifetimes (smallest eigenvalues) to encode a qubit.  

So how much advantage can one gain if the separation between atoms is small?
In Fig.~2(c) the longest lifetime possible is $109$ times longer than the
lifetime of the state of an isolated atom when the separation between adjacent
atoms is just $\lambda_0/4$. However, the second longest lifetime possible in
this system is just $4.5$ times longer than the lifetime of the state of an
isolated atom. Thus if we have chosen to construct a qubit from these two
states, our effective advantage would be just the lifetime of the second
state. The situation is a little better if we consider configurations that
have more symmetry: for example, consider the square in Fig. 2(b). Here the lifetimes for
the two collective states are comparable when the separation between atoms is
small. When the side of the square is equal to $\lambda_0/4$, the two largest
lifetimes are $4.6$ and $5.1$  times longer than the lifetime of the state of
an isolated atom.  
\begin{figure}[htb] 
\includegraphics[
       height=7 cm, keepaspectratio=true]{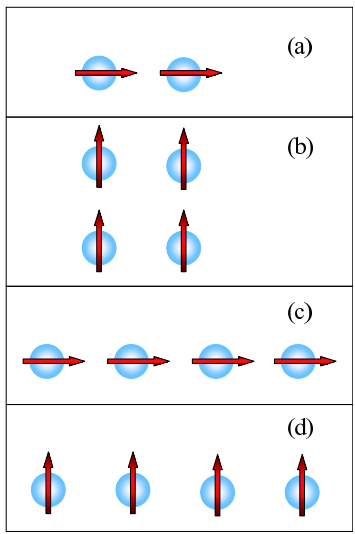}
\caption{   (a) The system of two two-level atoms in a line with each dipole
  moment co-aligned with the axis joining the atoms. (b) The system of four
  two-level atoms that lie on the corners of the square with each dipole
  moment co-aligned with one specific side of the square. (c) The system of
  four two-level atoms in a line with each dipole moment co-aligned with the
  axis connecting the atoms (d) The system of four two-level atoms in a line
  with each dipole moment perpendicular to the axis connecting the atoms.}
        \label{atomsConfig}
\end{figure}
\begin{figure}[htb] 
\includegraphics[
     height=10 cm, keepaspectratio=true]{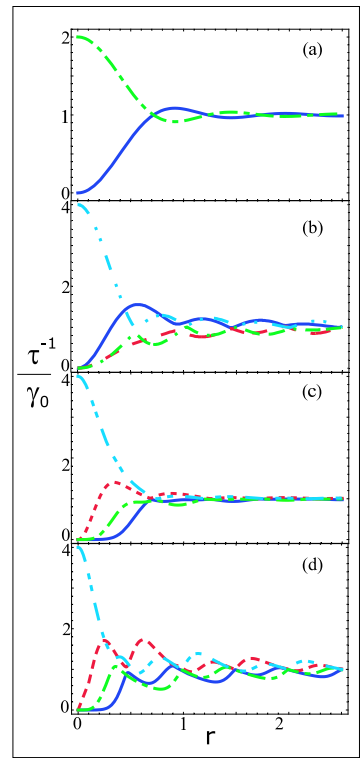}
\caption{   Inverse lifetimes (relative to $\gamma_0$) for configurations
  shown in Fig.~1. The two plots in (a) correspond to the configuration of two
  atoms in fig.~1(a). The dash-dot line corresponds to the symmetric state and
  the solid line to the antisymmetric state. The plots in (b)-(d) correspond
  to the configurations in fig. 1(b)-(d) respectively and the four lines
  correspond to the collective states that arise when one atom is in the
  excited state and all other atoms are in the ground state. }
        \label{gAtoms}
\end{figure}

We have shown how lifetimes for quantum states increase when the separation between atoms is small for a collection of a few atoms. Similar effect can be observed in a large collection of atoms. 
Spontaneous decay rate from a collection of a large number of two-level atoms was recently studied
in two configurations:
(a) $N$ atoms in a linear chain with the same distance between neighboring
atoms ~\cite{Ham05, Clem} and (b) a circular configuration, in which atoms
were placed equidistantly on the circumference of the circle~\cite{Ham05}.  

For the case of the linear chain, a numerical study showed that for large
number of atoms (40-100)~\cite{Ham05}, significant decay rate suppression was
observed when neighboring atoms were separated by a distance
$d<\lambda_0/2$. In the limit of $N\rightarrow\infty$, the decay rate was seen
to have a jump-like behavior at $\lambda_0/2$. This observation implied that
the spontaneous decay rate is unstable and susceptible to environmental
perturbations in the vicinity of this critical value.  

Analysis of the circular configuration of atoms offered more surprising
results. It was observed that for a given radius $r$ of the circular
configuration, once a certain critical number of atoms is reached the lifetime
of the maximally subradiant state increases exponentially and is now stable to
perturbations. For example, for $r=\lambda_0$, the critical number of atoms is
fourteen, and, at this moment, the smallest interatomic distance between atoms
is $0.45\times\lambda_0$.  

\section{Discussion}

Proposition \ref{prop:noMDFS} states under which conditions a DFS does not
exist, namely whenever the set of atoms is away from the Dicke limit. To find a DFS outside of the Dicke limit 
we therefore have to find systems that violate these 
conditions. One way
would be to consider a single-particle DFS with many such particles in the
system. Another way is to engineer a reservoir that is not
of homogeneous isotropic Markovian type.

Although the single-particle DFS is excluded for two-level atoms,
a DFS can be created by a collection
of particles that have a nontrivial internal structure. Unfortunately the
spherical symmetry of atoms induces selection rules for
angular momentum eigenstates that prohibit
the existence
of a DFS in isolated atoms. However, instead of atoms one can use molecules
that do not posses a spherical symmetry. A DFS in molecules is known in the
literature as spontaneous emission cancellation. It has been reported for
Rydberg states in Na$_2$~\cite{Xia} but the observation has been
questioned~\cite{Li}.

The conclusions we have reached about the existence of DFS in this work depend in a
fundamental way on the fact that the quantum system was embedded in
three-dimensional space. Requirements for existence of DFS in one-dimensional
structures are different. The logic behind Proposition~\ref{prop:noMDFS} tells us that in this case the separation between
atoms is no longer required to be zero for the existence of a DFS. (In one-dimensional structures, the reservoir transformation functions are reduced to scalars and condition for linear dependence is always satisfied.)
Therefore, a  multi-particle DFS may exist for atoms placed in an effectively
one-dimensional structure such as a waveguide or an optical
fiber. P. Zanardi and F. Rossi have discussed similar results in the context of semiconductor structures \cite{Zan99}.

In summary, we have developed a new way to determine the existence of both
single-particle and multi-particle DFS, for realistic systems described by a
Markovian master equation. We have demonstrated that a perfect multi-particle
DFS requires co-located particles (i.e.,~the Dicke limit) for systems placed in
a homogeneous isotropic Markovian reservoir, which is not possible.  Our
analysis shows however that it is nevertheless possible to have a  single-particle DFS involving many atoms. Also a
multi-particle DFS for atoms situated in one-dimensional  structures might
be possible. Furthermore we have established here a foundation for studying approximate
DFS and applied it to a set of $N$ two-level atoms interacting with a multi-mode radiation field.

\acknowledgments
R. I. K. acknowledges valuable discussions with F. A. Gr\"{u}nbaum.
This project has been funded by iCORE, CIAR, 
NSERC. R.I.K. and K.B.W. thank the NSF for financial support under ITR Grant
No. EIA-0205641, and the Defense Advanced Research Projects
Agency (DARPA) and the Air Force Laboratory, Air Force
Material Command, USAF, under Contract No. F30602-01-
2-0524.
 
\appendix
\section{Comparing two definitions for DFS}
\label{app:defDFS}
Here we show under which condition different conditions for DFS
are equivalent for systems coupled to a homogeneous isotropic
Markovian reservoir.
\begin{Proposition}
\label{prop:DFSequiv1}
        For a pure state $\rho=|\psi\rangle\langle\psi|$ of a system
        coupled to a  homogeneous isotropic Markovian reservoir,
        if $-i[\hat{\Delta},\rho]+L_D[\rho]=0$ then
        $L_D[\rho]=0$.
\end{Proposition}
\begin{proof}
Suppose that $-i[\hat{\Delta},\rho]+L_D[\rho]=0$ holds.
We then can evaluate the expression 
\begin{equation}
        0=\langle\psi|-i[\hat{\Delta},\rho]+L_D[\rho]|\psi\rangle=
        \langle\psi|L_D[\rho]|\psi\rangle.
\end{equation}
But
\begin{equation} 
  0 = \langle\psi|\, L_D\left[\rho \right]\, |\psi\rangle =
   \sum_{l=1}^M \lambda_l \left ( \left|\langle\psi|\hat{J}_l
          |\psi\rangle\right|^2 - \left\|\hat{J}_l |\psi\rangle\right\|^2  \right ).
\label{eq:dfs0}
\end{equation} 
$\hat{J}_l |\psi\rangle$ can generally be written as
$\hat{J}_l|\psi\rangle=c_l|\psi\rangle+|\psi_l^\bot\rangle$
with $|\psi_l^\bot\rangle$ some (non-normalized) state 
that is orthogonal to the state $|\psi\rangle$.
If we substitute this into Eq.~(\ref{eq:dfs0}) we get 
   $     \sum_{l=1}^M \lambda_l  \langle\psi_l^\bot|\psi_l^\bot\rangle=0$.
Because $\lambda_l>0$ for $l=1,\ldots, M$ we find
$\left\||\psi_\bot\rangle_l\right\|=0$, i.e., $\hat{J}_l|\psi\rangle=c_l|\psi\rangle$.
Thus $|\psi\rangle$ is an eigenstate of all the error generators $\hat{J}_l$ appearing in Eq.~\ref{eq:LD}.  As we have argued before for  the systems immersed in the homogeneous isotropic Markovian reservoir, operators $\hat{J}_l$ can have only zero eigenvalues. Using Proposition~\ref{prop:DFS0} we can infer that $L_D[\rho]=0$.
\end{proof}

\section{Fourier transform of the correlation Function}
\label{app:Fourier}
We derive Eq.~(\ref{eq:Gnm}) starting with Eq.~(\ref{Eq:cf}).
Expanding $G_{nm}^{ij}(\tau)$ in terms of $|\psi_{r \mathbf{k}}\rangle$ and introducing the unitary  operators
$U_{\mathbf{x}_n} = \exp (i\mathbf{x}_n\cdot\hat{\mathbf{P}})$ 
results in 
\begin{align}
\label{green}
        G_{nm}^{ij}(\tau)=& \theta(\tau)\sum_{r, s} \int \text{d}^3\mathbf{k}'\int \text{d}^3\mathbf{k}\;\;
         p_{s \mathbf{k}'} \nonumber \\ &\times \left \langle \psi_{s \mathbf{k}'}\left|
         \hat{E^i}(\mathbf{x}_n,\tau)\right|\psi_{r \mathbf{k}}\right\rangle \nonumber \\& \times
         \left\langle\psi_{r \mathbf{k}}\left| \hat{E^j}(\mathbf{x}_m,0)\right |\psi_{s \mathbf{k}'}\right\rangle
          \nonumber  \\ =&  \theta(\tau)\sum_{r, s}\int \text{d}^3\mathbf{k}\; 
          \int \text{d}^3\mathbf{k}'\; p_{s \mathbf{k}'}\nonumber \\&\times
          \left\langle \psi_{s \mathbf{k}'} \left|U_\text{R}^\dagger U_{\mathbf{x}_n}^\dagger
          \hat{E}^i(0)U_{\mathbf{x}_n}U_\text{R} \right| \psi_{r \mathbf{k}}\right\rangle \nonumber \\&\times
          \left\langle\psi_{r \mathbf{k}}\left|U_{\mathbf{x}_m}^\dagger \hat{E}^j(0)
          U_{\mathbf{x}_m}\right|\psi_{s \mathbf{k}'}\right\rangle 
          \nonumber \\  =&\theta(\tau)\sum_{r, s} \int \text{d}^3\mathbf{k}'\int \text{d}^3\mathbf{k}\;\;
           p_{s \mathbf{k}'}\text{e}^{i(\mathbf{k}-\mathbf{k}')\cdot(\mathbf{x}_n-\mathbf{x}_m)}\nonumber \\& \times
           \text{e}^{-i\tau \Omega_{r \mathbf{k} s \mathbf{k}'}} 
           \left \langle \psi_{s \mathbf{k}'} \left|\hat{E^i}(0)\right|\psi_{r \mathbf{k}}\right\rangle\;\nonumber \\& \times
           \left\langle\psi_{r \mathbf{k}}\left|\hat{E^j}(0)\right|\psi_{s \mathbf{k}'}\right\rangle.
\end{align}
We want to calculate the expression for 
\begin{equation}
        \widetilde{G}_{nm}^{ij}(\omega)\equiv\int_{-\infty}^\infty \text{d}\tau
        G_{nm}^{ij}(\tau)\exp(i\omega\tau).
\label{Eq:ft}
\end{equation}
The time-dependent part of the Green's function $G_{nm}^{ij}(\tau)$ in Eq.~(\ref{green}) is of the form $ \text{e}^{-i\tau \Omega_{r \mathbf{k} s \mathbf{k}'}} $. Thus the integral with respect to $\tau$ in the expression for $\widetilde{G}_{nm}^{ij}(\omega)$ is  
\begin{align}
 & \lim_{\eta\rightarrow 0}\int_0^\infty \! \!\!\text{d}\tau\,
  \text{e}^{i\tau(\omega-\Omega_{r \mathbf{k} s \mathbf{k}'})-\eta\tau} \nonumber \\ 
 &=\lim_{\eta\rightarrow 0} 
  \frac{i}{\omega-\Omega_{r \mathbf{k} s \mathbf{k}'}+i\eta} \nonumber \\&=
 \left(\pi \delta(\omega -\Omega_{r \mathbf{k} s \mathbf{k}'})
  +i\frac{\cal P}{\omega-\Omega_{r \mathbf{k} s \mathbf{k}'}}\right).
\end{align}
Then the expression for the Fourier transform of the Green's function is 
\begin{align}
 \widetilde{G}_{nm}^{ij}(\omega)=&\sum_{r,s}\int \text{d}^3\mathbf{k}' \; \int \text{d}^3\mathbf{k} \; p_{s \mathbf{k}'}\,
  \text{e}^{i(\mathbf{k}-\mathbf{k}')\cdot(\mathbf{x}_n-\mathbf{x}_m)} \nonumber \\&\times
  \langle\psi_{s \mathbf{k}'}|\hat{E^i}(0)|\psi_{r \mathbf{k}}\rangle \; 
  \langle \psi_{r \mathbf{k}}|\hat{E^j}(0)| \psi_{s \mathbf{k}'}\rangle  
\nonumber\\ & \times
  \left(\pi \delta(\omega -\Omega_{r \mathbf{k} s \mathbf{k}'})
  +i\frac{\cal P}{\omega-\Omega_{r \mathbf{k} s \mathbf{k}'}}\right).
\end{align}
We use this expression in Eq.~(\ref{eq:Gnm}).

\section{Linear independence}
\label{app:liIn}
Here we verify the claim made in the proof of Proposition \ref{prop:noMDFS}
that the set of functions
$\{P_j(\mathbf{k})\text{e}^{i\mathbf{k}\cdot\mathbf{x}_n}\}_{j=1}^N$, where
$P_j(\mathbf{k})$ are polynomials in Cartesian components of $\mathbf{k}$ defined on the sphere of radius $r$ ($|\mathbf{k}|=r$), are linearly independent if $\mathbf{x}_n\neq \mathbf{x}_m$ for $n\neq m$.

Before we proceed, we need to introduce some notation.
Let components of the vector $\mathbf{k}$ be denoted by $k_1,\,k_2,\,k_3$ and let $\mathbf{\alpha}=(\alpha_1,\,\alpha_2,\,\alpha_3)$  be a collection of non-negative integers. We define $\mathbf{k}^\mathbf{\alpha}\equiv k_1^{\alpha_1} k_2^{\alpha_2} k_3^{\alpha_3}$. Then a polynomial $P(\mathbf{k})$ is denoted as $P(\mathbf{k})=\sum_\alpha a_\alpha \mathbf{k}^\alpha$, where coefficients $a_\alpha$ are non-zero only for finite collection of $\alpha$. Let $|\alpha|\equiv \alpha_1+\alpha_2+\alpha_3$.
The \emph{full degree of the polynomial} $P(\mathbf{k})$, denoted by $\text{deg}(P)$, is the largest $|\alpha|$ of all terms in $P(\mathbf{k})$ with non-zero coefficients $a_\alpha$.

\begin{Proposition}
\label{prop:linInd}
Let $P_1(\mathbf{k}),\,P_2(\mathbf{k}),\,\cdots,\,P_N(\mathbf{k})$ be non-zero
polynomials in Cartesian components of vector $\mathbf{k}$. Then the set of functions
$\{P_1(\mathbf{k})e^{i\mathbf{k}\cdot\mathbf{x}_1},\,P_2(\mathbf{k})e^{i\mathbf{k}\cdot\mathbf{x}_2},\,\cdots,\,P_N(\mathbf{k})e^{i\mathbf{k}\cdot\mathbf{x}_N}\}$
is linearly independent when vectors with real
entries $\mathbf{x}_1, \mathbf{x}_2, \cdots, \mathbf{x}_N$ are distinct.
\end{Proposition}
\begin{proof}
Functions $P_j(\mathbf{k})\text{e}^{i\mathbf{k}\cdot\mathbf{x}_n}$ $j=1,\ldots, N$ are linearly independent, if they cannot be expressed in the form
\begin{align} 
c_1 P_1(\mathbf{k})e^{i\mathbf{k}\cdot\mathbf{x}_1}+c_2 P_2(\mathbf{k})e^{i\mathbf{k}\cdot\mathbf{x}_2}+\cdots \nonumber \\+c_N P_N(\mathbf{k})e^{i\mathbf{k}\cdot\mathbf{x}_N}=0
\label{eq:linInd}
\end{align}
with $c_j$ ($j=1,\ldots, N$) constants not all equal to zero.

We assume that~Eq. (\ref{eq:linInd}) holds. Then we use mathematical induction to show that all $c_i$'s are 0.
  
For $N=1$, we have   $ c_1 P_1(\mathbf{k})e^{i\mathbf{k}\cdot\mathbf{x}_1}=0$. Since $P_1(\mathbf{k})e^{i\mathbf{k}\cdot\mathbf{x}_1}$ is a non-zero function, $c_1=0$.

Now suppose the statement is true for $N$ functions, i.e. if  
\begin{align} 
c_1 P_1(\mathbf{k})e^{i\mathbf{k}\cdot\mathbf{x}_1}+c_2 P_2(\mathbf{k})e^{i\mathbf{k}\cdot\mathbf{x}_2}+\cdots \nonumber \\+c_N P_N(\mathbf{k})e^{i\mathbf{k}\cdot\mathbf{x}_N}=0,
\end{align}
then $c_i=0$ for $1\leq i\leq N$.

We want to show that if 
\begin{align}
c_1 P_1(\mathbf{k})e^{i\mathbf{k}\cdot\mathbf{x}_1}+c_2 P_2(\mathbf{k})e^{i\mathbf{k}\cdot\mathbf{x}_2}+\cdots\nonumber \\+c_{N+1} P_{N+1}(\mathbf{k})e^{i\mathbf{k}\cdot\mathbf{x}_{N+1}}=0,
\label{eq:lcN+1}
\end{align}
then $c_i=0$ for $1\leq i\leq N+1$.

Let $\mathbf{x}_n=(\mathbf{x}_{n,1},\,\mathbf{x}_{n,2},\,\mathbf{x}_{n,3})$ and let $m
_n=\text{deg}(P_n)+1$. Consider the linear differential operator $L_n=\sum_{j=1}^3 (\frac{\partial}{\partial k_j} - i\mathbf{x}_{n,j} \mathbf{1})^{m_n}$ and observe that 
\begin{equation}
 L_n\left(P_n(\mathbf{k})e^{i\mathbf{k}\cdot\mathbf{x}_n}\right)=0.
\label{eq:con1}
\end{equation}
\begin{equation}
\text{ For } l\neq n,\,\,  L_n\left(P_l(\mathbf{k})e^{i\mathbf{k}\cdot\mathbf{x}_l}\right)=Q_{l, n}(\mathbf{k})e^{i\mathbf{k}\cdot\mathbf{x}_l},
\label{eq:con2}
\end{equation}
where $Q_{l,n}(\mathbf{k})$ is a non-zero polynomial with $\text{deg}(P_l)=\text{deg}(Q_{l,n})$.

The statement (\ref{eq:con1}) holds because the power of the term $(\frac{\partial}{\partial k_j} - i\mathbf{x}_{n,j} \mathbf{1})$ in $L_n$ exceeds the degree of variables $k_1,\,k_2,\,k_3$ in polynomial $P_n(\mathbf{k})$ at least by 1. The statement (\ref{eq:con2}) is verified later.

We apply a linear operator $L_{N+1}$ to the Eq.~(\ref{eq:lcN+1}). 
Then
\begin{align}
c_1 Q_{1, N+1}(\mathbf{k})e^{i\mathbf{k}\cdot\mathbf{x}_1}+c_2 Q_{2,N+1}(\mathbf{k})e^{i\mathbf{k}\cdot\mathbf{x}_2}+\cdots\nonumber\\+c_{N} Q_{N, N+1}(\mathbf{k})e^{i\mathbf{k}\cdot\mathbf{x}_{N}}=0.
\end{align}
Polynomial $Q_{l,N+1}(\mathbf{k})$ is non-zero  and it has the same degree as $P_l(\mathbf{k})$ for $1\leq l\leq N$. Therefore by induction hypothesis $c_1=c_2=\cdots=c_N=0$. Thus $c_{N+1} P_{N+1}(\mathbf{k})e^{i\mathbf{k}\cdot\mathbf{x}_{N+1}}=0$, which is only possible when $c_{N+1}=0$.

In order to check that statement (\ref{eq:con2}) is correct, we introduce polynomials $Q_{l,n}^j(\mathbf{k})$ such that  \begin{equation}
Q_{l,n}^j(\mathbf{k})e^{i\mathbf{k}\cdot\mathbf{x}_l}=\left(\frac{\partial}{\partial k_j} - i\mathbf{x}_{n,j} \mathbf{1}\right)^{m_n}P_l(\mathbf{k})e^{i\mathbf{k}\cdot\mathbf{x}_l}.
\end{equation}
Notice that $Q_{l,n}(\mathbf{k})=\sum_{j=1}^3 Q_{l,n}^j(\mathbf{k})$. Since $\mathbf{x}_l\neq\mathbf{x}_n$, there exists $j\in\{1,\,2,\,3\}$ so that $\mathbf{x}_{l,j}\neq\mathbf{x}_{n,j}$. We claim that $\text{deg}(Q_{l,n}^j)=\text{deg}(P_l)$.
Indeed for $m_n=1$,
\begin{equation}
Q_{l,n}^j(\mathbf{k})= \left(\frac{\partial P_l}{\partial k_j}(\mathbf{k})+i(\mathbf{x}_{l,j}-\mathbf{x}_{n,j})P_l(\mathbf{k})\right).
\end{equation}

Therefore $Q_{l,n}^j(\mathbf{k})$ has the same total degree as $P_l(\mathbf{k})$. Suppose the claim is correct for $m_n=M$. Then for $m_n=M+1$, 
\begin{align}
Q_{l,n}^j(\mathbf{k})e^{i\mathbf{k}\cdot\mathbf{x}_l} &= \left(\frac{\partial}{\partial k_j} - i\mathbf{x}_{n,j} \mathbf{1}\right)^{M}\nonumber \\ & \times\left(\frac{\partial P_l}{\partial k_j}(\mathbf{k})+i(\mathbf{x}_{l,j}-\mathbf{x}_{n,j})P_l(\mathbf{k})\right)e^{i\mathbf{k}\cdot\mathbf{x}_l}.
\end{align}
By assumption $\text{deg}(Q_{l,n}^j(\mathbf{k}))=\text{deg}(\frac{\partial P_l}{\partial k_j}(\mathbf{k})+i(\mathbf{x}_{l,j}-\mathbf{x}_{n,j})P_l(\mathbf{k}))$ and $\text{deg}(\frac{\partial P_l}{\partial k_j}(\mathbf{k})+i(\mathbf{x}_{l,j}-\mathbf{x}_{n,j})P_l(\mathbf{k}))=\text{deg}(P_l(\mathbf{k})$).

Also if polynomial $P_l(\mathbf{k})$ is  constant, i.e.   $P_l(\mathbf{k})=c\neq0$, then polynomial $Q_{l,n}(\mathbf{k})$ has the form $Q_{l,n}(\mathbf{k})=c\sum_{j=1}^3 \left(i(\mathbf{x}_{l,j}-\mathbf{x}_{n,j})\right)^{m_n}$. Since $\mathbf{x}_l\neq\mathbf{x}_n$ for $l\neq n$, polynomial $Q_{l,n}(\mathbf{k})$ is also non-zero.
\end{proof}

\section{Relationship between decoherence operator and reduced matrix.}
\label{app:reducedmatrix}
We analyze here the relationship between the reduced matrix $\overleftrightarrow{\gamma}$ and the decoherence operator $\hat{\Gamma}$. In the Propositions below  we demonstrate the relationship between the eigenvalues of the matrix $\overleftrightarrow{\gamma}$ and the decoherence matrix $\hat{\Gamma}$, as well as establish the connection between eigenvectors for these two matrices.

\begin{Proposition}
\label{eigv}
If $\lambda$ is an eigenvalue of $\overleftrightarrow{\gamma}$, then $\gamma_0 \lambda$ is an eigenvalue of decoherence matrix $\hat{\Gamma}$. 
\end{Proposition}
\begin{proof}

Let $\left(x_1\cdots x_N\right)^\text{T}$  be an eigenvector of matrix $\overleftrightarrow{\gamma}$. Then  $\overleftrightarrow{\gamma} \left(x_1\cdots x_N\right)^\text{T}=\lambda  \left(x_1\cdots x_N\right)^\text{T}$ for some real number $\lambda$. In terms of components this can be written as $\sum_{k=1}^N \gamma_{jk} x_k=\lambda x_j$. 

We would like to show that $\gamma_0\lambda$ is an eigenvalue for operator $\hat{\Gamma}$. This will be true iff there exists state $|\psi\rangle$ such that $\langle\psi|(\hat{\Gamma}/\gamma_0-\lambda I)|\psi\rangle=0$. Consider a normalized state $|\psi\rangle =\sum_{l=1}^N \,x_l |1,l\rangle$, where the state $|1,l\rangle$ corresponds to $l^{\text{th}}$ atom in the excited state, while all other atoms are in their ground state. The state $|0\rangle$ denotes the collective ground state for all $N$ atoms. Then $\hat{\sigma}_k |\psi\rangle=\delta_{kl}x_l |0\rangle$ and $\langle\psi|\hat{\sigma}_j^\dagger \hat{\sigma}_k|\psi\rangle=x_j x_k$. Thus 
\begin{align}
        \langle\psi|(\hat{\Gamma}/\gamma_0-\lambda I)|\psi\rangle= &\langle\psi|(\sum_{j,k=1}^N \gamma_{jk} 
        \hat{\sigma}_j^\dagger \hat{\sigma}_k)|\psi\rangle-\lambda \nonumber \\ =&\sum_{j,k=1}^N 
        \gamma_{jk}\langle\psi|\hat{\sigma}_j^\dagger \hat{\sigma}_k|\psi\rangle -\lambda\nonumber \\ =&
        \sum_{j,k=1}^N \gamma_{jk}x_j x_k-\lambda=0
\end{align}
\end{proof}

\begin{Proposition}
Eigenvalues of the matrix $\gamma_0\overleftrightarrow{\gamma}$ are inverse lifetimes for collective states of the system with $N-1$ two-level atoms in the ground state and one atom in the excited state.
\end{Proposition}
\begin{proof}
Lifetimes for $N$ asymptotically separated atoms is the same as for $N$ isolated atoms. Thus inverse lifetimes for collective states with one excited atom (all other atoms in the ground state) are equal to $\gamma_0$. There will be $N$ such states. Matrix $\overleftrightarrow{\gamma}$ has $N$ eigenvalues and for infinite separation between atoms its eigenvalues are $1$.
\end{proof}

\begin{Proposition}
If $\lambda$ is an eigenvalue of $\overleftrightarrow{\gamma}$, then $(N-2+\lambda)\gamma_0$ is an eigenvalue of the decoherence operator $\hat{\Gamma}$ in the system of $N$ two-level atoms.
\end{Proposition}
\begin{proof}
We proceed as in the proof for Proposition~\ref{eigv}. Let $\left(x_1\cdots x_N\right)^\text{T}$  be such that  $\sum_{k=1}^N \gamma_{jk} x_k=\lambda x_j$. 

Consider the normalized state $|\psi\rangle =\sum_{l=1}^N \,x_l |0,l\rangle$, where state $|0,l\rangle$ corresponds to $l^\text{th}$ atom in the ground state, while all other atoms are in their excited state. The state $|1\rangle$ denotes the collective excited state for all $N$ atoms. Then $\hat{\sigma}_k^\dagger |\psi\rangle=\delta_{kl}x_l |1\rangle$,  $\langle\psi|\hat{\sigma}_j^\dagger \hat{\sigma}_k|\psi\rangle=x_j x_k$ for $k\neq j$ and $\langle\psi|\hat{\sigma}_k^\dagger \hat{\sigma}_k|\psi\rangle=\sum_{j\neq k=1}^N x_j^2=1-x_k^2$. Thus 
\begin{align}
&\langle\psi|\left(\hat{\Gamma}/\gamma_0-(N-2+\lambda) I\right)|\psi\rangle \nonumber \\ =&\sum_{j,k=1}^N \gamma_{jk}\langle\psi|\hat{\sigma}_j^\dagger \hat{\sigma}_k|\psi\rangle -(N-2+\lambda) \nonumber \\ =& \sum_{j,k=1}^N \gamma_{jk} x_j x_k+N-2\sum_{k=1}^N x_k^2-(N-2+\lambda)=0
\end{align}
\end{proof}
We have shown how knowledge of the reduced decoherence matrix $\overleftrightarrow{\gamma}$ allows us to determine $2N$ eigenvalues of the full decoherence operator $\hat{\Gamma}$. We also were able to derive the expressions for the corresponding eigenstates, which are $|\psi_n\rangle =\sum_{l=1}^N \,x_l^n |0,l\rangle$ and $|\phi_n\rangle =\sum_{l=1}^N \,x_l^n |1,l\rangle$ with $n=1,\ldots, N$, where $|1,l\rangle$ is the state with only $l^\text{th}$ atom in excited state, $|0,l\rangle$ is the state with only $l^\text{th}$ atom in the ground state, and $(x_1^n \ldots x_N^n)^T$ is the $n^\text{th}$ eigenstate of matrix  $\overleftrightarrow{\gamma}$.

\end{document}